\documentclass[11pt,nofootinbib]{article}
\usepackage{a4wide}
\usepackage{amsfonts}
\usepackage{amssymb}
\usepackage{amsmath}
\usepackage{graphicx}
\usepackage{booktabs}
\usepackage{array}
\usepackage{color}
\usepackage{ulem}
\usepackage[small,bf]{caption}
\usepackage{cite}
\usepackage[usenames,dvipsnames]{xcolor}
\usepackage{subfigure}
\usepackage{verbatim}
\usepackage{hyperref}

\newcommand{\mdm}{M_{\text{DM}}}

\newcommand{\rhodm}{\rho_{\text{DM}}}

\newcommand{\avdm}{\bar{v}_{\text{DM}}}

\newcommand{\vesc}{v_{\text{esc}}}
\newcommand{\sigmadmA}{\sigma_{\text{DM}-N}}
\DeclareMathOperator{\diff}{\text{d}}

\numberwithin{equation}{section}
\begin{document}
\begin{titlepage}
		
\begin{center}
{ \bf\LARGE Dark Matter Induced Brownian Motion} 
\\[8mm]
Ting Cheng$^{\, a}$ \footnote{E-mail: \texttt{ting.cheng@mpi-hd.mpg.de}, now at Max-Planck-Institut f\"ur Kernphysik, Heidelberg, Germany},
Reinard Primulando$^{\, b}$ \footnote{E-mail: \texttt{rprimulando@unpar.ac.id}},
Martin Spinrath$^{\, a}$ \footnote{E-mail: \texttt{spinrath@phys.nthu.edu.tw}}
\\[1mm]
\end{center}
\vspace*{0.50cm}
\centerline{$^{a}$ \it Department of Physics, National Tsing Hua University, Hsinchu 30013, Taiwan}
\vspace*{0.2cm}
\centerline{$^{b}$ \it Center for Theoretical Physics, Department of Physics, Parahyangan Catholic University,}
\centerline{\it Jl. Ciumbuleuit 94, Bandung 40141, Indonesia}
\vspace*{0.2cm}
		
\vspace*{1.20cm}
		
\begin{abstract}
\noindent
We discuss a novel approach for directional, light dark matter searches inspired
by the high precision position measurements achieved in gravitational wave detectors.
If dark matter interacts with ordinary matter, movable masses are subject to
an effect similar to Brownian motion induced by the scattering with
dark matter particles which exhibits certain characteristics and could be observed. We
provide estimates for the sensitivity of a hypothetical
experiment looking for that motion. Interestingly, if successful, our approach would allow to
constrain the local distribution of dark matter momentum.			
\end{abstract}

\end{titlepage}

\setcounter{footnote}{0}

\section{Introduction}

Cosmological and astrophysical data provides
overwhelming evidence for dark matter (DM). Unfortunately,
this data does not tell us anything definite about the nature of 
DM itself.

To solve this riddle, there has been tremendous experimental effort
in the last decades focussing, in particular, on theoretically
well-motivated weakly interacting TeV scale DM, so called
Weakly Interacting Massive Particles (WIMPs).
These efforts remain until today without any conclusive evidence for
a discovery.
For that reason, recently other potential mass regions for DM are more
seriously considered and new ideas are developed to test them
experimentally, see, e.g.~\cite{Alexander:2016aln}.
In this paper we follow this line and discuss a novel experimental approach
for light DM. As we will see our method would work, in principle,
to very small masses. However, under more realistic assumptions, 
its sensitivity lies in the range above about 10~MeV/c$^2$.

Our proposal is motivated by the great achievements in laser interferometry
for gravitational wave detectors, but as we will see later LIGO and other
current earth-bound gravitational wave detectors are not well suited for our method.

We are not the first to propose to use gravitational wave detectors or interferometers in general as DM detectors, 
see for instance~\cite{Riedel:2012ur, PhysRevLett.114.161301, Arvanitaki:2015iga, Stadnik:2015xbn, Branca:2016rez, Riedel:2016acj, Jung:2017flg, Pierce:2018xmy, Morisaki:2018htj, Grote:2019uvn}. Nevertheless, our approach is
very different from theirs. They usually focus on very light
DM with masses well below 1~keV/c$^2$, where DM behaves more like a classical field with
a very long wave-length. In our method the particle nature of DM is essential.
In fact, it is loosely 
inspired by the work of one of the authors~\cite{Domcke:2017aqj}.
Interestingly, results of the KWISP detector were presented which
is looking for dark energy particles with an opto-mechanical setup~\cite{Cuendis:2019mgz}
somewhat similar to our proposal. There has also been a search for
very heavy DM using displacement sensors \cite{Kawasaki:2018xak}. 

The idea of this paper is based on the picture that any movable target with mass $M_T$,
is actually in a bath of DM.
If DM has some interaction with ordinary matter, that
will induce some random motion of the target, which is conceptually similar to Brownian motion.
We call this here Dark Brownian Motion (DBM) which can in principle be observed.
It is experimentally easier to study the DBM if the target is constrained to a certain volume, e.g., if it is a pendulum.
The concrete system is not relevant here, but
we assume that the
motion of the target in absence of any external forces can be understood well.
In the following section, we will discuss the details of our method.

\section{The Detection Method}

Since the earth has a relative velocity to the DM bath we expect that the DBM
has a preferred direction. This directional dependence is well-known for directional
dark matter detection with more conventional detectors~\cite{Spergel:1987kx}. In particular,
we adopt here the formalism for the DM recoil momentum spectrum as it was discussed,
for instance, in~\cite{Gondolo:2002np}. To keep the discussion simple, we will assume
that the DM
interaction with ordinary matter is fully elastic. The total event rate, $R$, is given
by
\begin{align}
	\label{eq:Rate}
	R &= (Z+N)^2 \, \sigmadmA \, \frac{M_T}{M_{\text{mol}} } \, \frac{\rhodm}{\mdm} \,  \bar{v}_{\text{DM}} \; \nonumber \\
	&= 0.37 \left( \frac{{Z+N}}{12} \, \frac{\sigmadmA}{10^{-31}\, \text{cm}^2} \,
	\frac{M_T}{10^{-3} \, \text{g}} \,
	\frac{\rhodm}{0.3 \,\text{GeV/cm}^3}\,
	\frac{ 20 \,\text{MeV}}{\mdm} \,  
	\frac{\bar{v}_{\text{DM}}}{341 \,\text{km/s}} \;\right) \frac{1}{\text{s}} \;,
\end{align}
where 
$Z$ and $N$ are the atomic number and the number of neutrons respectively, $\sigmadmA$ is the average cross section of DM with a target nucleon,
$M_T$ the target mass, $M_{\text{mol}}$ is the molar mass of the target,
$\rhodm = 0.3$~GeV/cm$^3$ \cite{Smith:2006ym} is our assumed local DM energy density,
$\bar{v}_{\text{DM}} = \int \text{d}^3 v_{\text{DM}}  \,  v_{\text{DM}} f(\vec{v}_{\text{DM}})$
is the average local DM velocity. We would like to point out here already, that although
the interaction rate has units of inverse time it does not correspond to a well-defined frequency
since the hits occur at random intervals with random recoil momenta.
In fact, due to the lack of periodicity the Fourier transformation of the
DM hit signal usually has no dominant peak around the average hit rate, but
rather many random small peaks scattered over all frequencies. Therefore, we perform
our analysis in the time-domain and not in the frequency-domain.

We assume that DM couples dominantly
to nucleons in an isospin conserving manner,
and throughout
the paper we assume $^{12}$C as target material
motivated by the use of graphene as reflective material \cite{doi:10.1063/1.4795787}
and tunable oscillators \cite{GrapheneOscillator}
which might be able to give us a light, tunable mirror in a laboratory size interferometer.
Actually, the physics in this letter
is not related to (laser) interferometers. We only require a precise
measurement of the position and/or velocity of a macroscopic object.
To our knowledge laser interferometers are the best technology for that
purpose and therefore we will consider them as an example.
Other potential technologies could be, for instance,
magnetic resonance force microscopy, see \cite{Barabanov:2018eec}
and references therein or optically levitated
microspheres, cf.\ \cite{Kawasaki:2018xak}.
Therefore, we encourage the reader to be open-minded and consider our proposal
independently from that.

For the
DM velocity distribution we assume the Standard Halo Model (SHM)
\begin{equation}
f_{\mathrm{gal}}(\mathbf{v})=\left\{\begin{array}{ll}{\frac{N}{\left(2 \pi \sigma_{v}\right)^{3 / 2}} \exp \left(-\frac{|\mathbf{v}|^{2}}{2 \sigma_{v}^{2}}\right)} & {\text { if }|\mathbf{v}|<v_{\mathrm{esc}}} \;, \\ {0} & {\text { if }|\mathbf{v}|>v_{\mathrm{esc}}} \;,\end{array}\right.
\end{equation}
where $v_{\mathrm{esc}} = 550$~km/s, $\sigma_{v} = 220$~km/s~\cite{Smith:2006ym} and
for the translation between the galactic rest frame and the local laboratory frame we
used the formulas in Appendix~A of~\cite{Mayet:2016zxu}.

It is not clear that the SHM is a good description of the DM velocity distribution.
In a recent work~\cite{Bringmann:2018cvk}, for instance, it was argued that
some fraction of the DM could receive a significant shift towards larger velocities
opening up direct detection constraints to a lower mass region.
Similarly, the DM velocity profile does not need to be isotropic and there could be
streams of DM passing through the solar system, see, for instance, \cite{OHare:2018trr}.
A higher DM velocity and/or anisotropy could make our setup
more sensitive. Nevertheless, a detailed study of such possibilities goes beyond the scope
of the current work and to make our results more easily
comparable to other studies, we will follow the conventional SHM assumption.

Under these assumptions and the formalism of~\cite{Gondolo:2002np} we can
simulate the distribution of recoil momenta.
For the incoming momenta we picked random values following the SHM distribution. The recoil
momenta are then fully fixed by choosing random values for the scattering angles, $\theta$ and $\phi$, which we
assumed to have a flat distribution in $\cos \theta$ and $\phi$.
This allows us to evaluate the
asymmetry parameter
\begin{align}
A = \frac{N_+ - N_-}{N_+ + N_-} = p_+ - p_-\;,
\end{align}
where $N_{\pm}$ is the number of events of the DM hitting the detector with positive/negative target recoil
momentum $q_R$ with $|q_R| > q_{\text{min}}$ along the axis we are probing.
The parameter $q_{\text{min}}$ is a cutoff parameter which we will discuss
later in more detail. We have also introduced here the likelihoods $p_\pm = N_\pm/(N_+ + N_-)$.

To determine $A$ we simulated for each parameter point one million scattering events to
get the distribution of the recoil momenta, from which we can derive $p_\pm$ easily.
We estimate the statistical uncertainty for $A$ using ordinary
error propagation with the 1$\sigma$ uncertainty of the number of events given by
$\sigma_{\pm} = \sqrt{N_{\pm}}$,  
and find
\begin{equation}
\sigma_A^2 = \frac{4}{R \, \Delta t} \, p_+\, p_- 
\end{equation}
with $\Delta t$ as the length of data taking, which we have fixed 
for the rest of the paper to be 10~minutes.

At this point 
we want to sketch how the counting of events could
be performed. We mentioned already that we assume to have a
mathematical model for the experiment which predicts the position
and velocity of the target mass in the absence of forces.
Additionally, we also assume that the time resolution of the detector can resolve the DM hits.
Suppose now the data is taken in time bins $t_1$, $t_2$, $\ldots$
and the target is hit by a DM particle at some time between $t_h$ and $t_{h+1}$.  
The target position and velocity at $t_{h+1}$ will then differ from its
predicted values given by the values measured at the previous time bins, 
and from the
differences we can reconstruct the recoil momentum of the 
DM hit.
For a harmonic oscillator, this can be
done straight-forwardly using a Runge-Kutta algorithm as we convinced ourselves.
In fact, for the asymmetry we only really need to reconstruct the sign 
of the recoil momenta,
which is comparatively easy.

\begin{figure}
	\centering
	\includegraphics[width=0.65\textwidth]{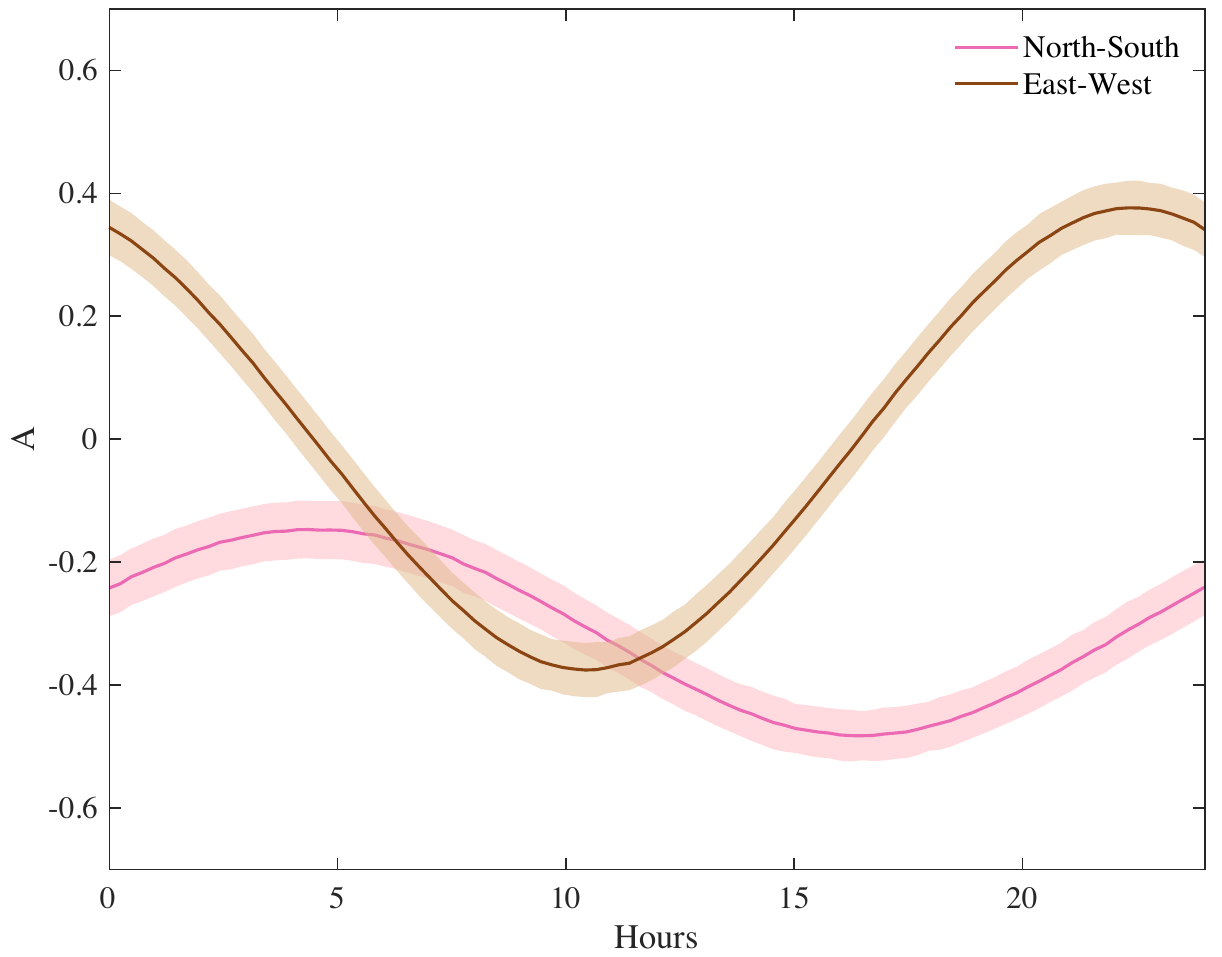}%
	\caption{
		The expected time dependence of the asymmetry $A$ in
		north-south (pink) and east-west direction (brown) in the earth frame within 24 hours
		in Hsinchu on June 1st.
		The colored bands represent the estimated 1$\sigma$ uncertainty.
		For more details see main text.
		\label{fig:AsymmetryTimeDependence}}
\end{figure}

While in this paper we will use the value of $A$ from 10 minutes 
of hypothetical 
data taking, it is interesting to see how the value of $A$ changes over time 
due to the change of the relative orientation of the detector. In Fig.~\ref{fig:AsymmetryTimeDependence} we plot the time dependence of the asymmetry
for an imaginary experiment located in Hsinchu within 24 hours on June 1st, 2020.
For the error bars
we have assumed that we take data for ten minutes with a target mass
of $10^{-3}$ g and $\sigmadmA = 10^{-31}$~cm$^2$.

Figure~\ref{fig:AsymmetryTimeDependence} also contains some information about the directional dependence of the asymmetry parameter. The north-south direction has the largest absolute value of $A$, and the east-west direction has a larger value of daily modulation of $A$. There is some room for optimization here by rotating the experiment, in terms of the absolute value or the daily modulation of the asymmetry. However for simplicity, in the rest of the paper, we choose the east-west alignment of our hypothetical experiment. Moreover we simulate the data at 10pm local Hsinchu time.
Then $A \approx 0.37$, $\avdm \approx 341$~km/s and
the relative velocity of the lab with respect to the DM halo along this direction
is $v_{\text{lab,EW}} \approx -183$~km/s. A positive $A$ here means that there are more hits
expected with a recoil momentum in the eastern direction.

We would like to clearly emphasize at this point,
that we propose here to count hits which then enter
in the determination of $A$. The hit rates in a certain
direction should be modulated with a frequency of roughly 1/day
and 1/year translating into a modulation of $A$, but we do
not propose to measure these frequencies
directly. One of the major reasons to consider, nevertheless,
$A$ as the relevant observable, and not just the total
number of events or the measured momentum distribution itself,
is its characteristic of a more pronounced time dependence,
which can be a powerful discrimination from backgrounds.
Moreover, it is more robust against uncertainties of 
the recoil momentum measurement after applying
a momentum cutoff as will be introduced later.

\section{Estimated Sensitivity}

In this section we will provide some simple
sensitivity estimates to illustrate the physics behind it
and what kind of precision and background
reduction would be needed to observe DBM and, in particular, use it as
a competitive constraint compared to more conventional
DM searches. 

\begin{figure}
	\centering
	\includegraphics[width=0.7\textwidth]{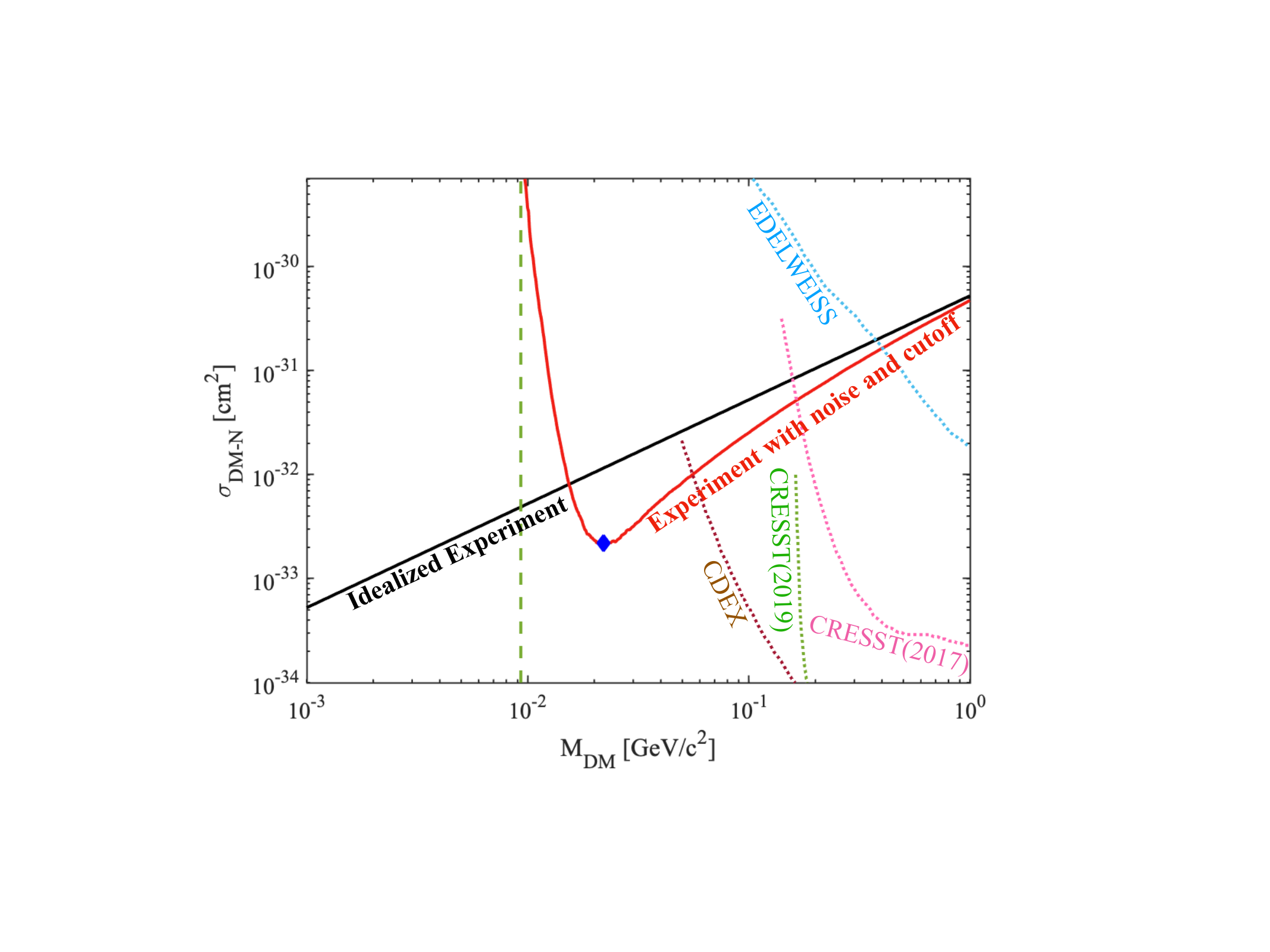}%
	\caption{
		The estimated 2$\sigma$ sensitivity of our experimental setup for a total target mass
		of 10$^{-3}$~g. The black line is for an idealized experiment
		and
		the red line includes our noise estimates a minimal momentum cutoff,
		$q_{\text{min}} = 2 \times 10^{-23}$~kg\,m/s.
		We also show the exclusion bounds from CRESST \cite{Angloher:2017sxg, Abdelhameed:2019hmk},
		EDELWEISS (Migdal) \cite{Armengaud:2019kfj}
		and CDEX (Migdal) \cite{Liu:2019kzq} as pink, green, light blue and dark red dotted
		lines, respectively.
		The blue diamond corresponds to the blue diamond in Fig.~\ref{fig:Avsq}.
		For more details, see main text.
		\label{fig:CrossSectionVsDMmass}}
\end{figure}

In Fig.~\ref{fig:CrossSectionVsDMmass} we show our estimate for the sensitivity.
We begin our discussion with an idealized experiment with infinitely precise
momentum measurement and no background events. For a total target mass
of $M_T = 10^{-3}$~g we find the black line as 2$\sigma$ exclusion level,
i.e.\ $\langle A \rangle/\sigma_A = 2$.
The dependence of this line can be easily understood.
For smaller masses, the DM flux, and hence the event rate, increases while the shape of
the recoil momentum distribution remains the same, i.e.
\begin{equation}
\frac{\langle A \rangle^2}{\sigma_A^2} = R \, \Delta t \, \frac{(p_+ - p_-)^2}{4\,  p_+\, p_-} \;.
\end{equation}
In this formula
the only explicit dependence on the DM mass and the DM-nucleon cross section
is in $R$, cf.~Eq.~\eqref{eq:Rate}.

In reality though there will be experimental uncertainties which alter
the sensitivity. To illustrate their potential impact we introduce two
sources of uncertainty.

The first is a generic background giving a recoil
momentum of the test masses. That could be seismic noise,
nearby traffic, radioactivity, etc.
Many of them can be reduced by choosing the location
and materials of the experiment,
e.g.~putting it underground, choosing low radioactivity materials, and
using a suspension system to reduce seismic noise.
Furthermore, as
already mentioned many of these backgrounds could be easily discriminated from
a potential DM signal due to the different characteristic time-dependencies of the asymmetries
which we do not explicitly consider in this section though.
Nevertheless, we want to discuss some estimates for two particle backgrounds
which are definitely present as examples.
Those are neutrinos and hits from residual air molecules.
The largest neutrino flux on earth is from the solar $pp$ neutrinos with
$\Phi_\nu \approx 6 \times 10^{10}$~cm$^{-2}\,$s$^{-1}$ at an energy of about 0.4~MeV
corresponding to a nucleon cross section to carbon of about $10^{-44}$~cm$^2$.
For our assumed target mass of $10^{-3}$~g this corresponds to a neutrino event
rate of $\mathcal{O}(10^{-14})$~s$^{-1}$. Obviously, even including all other neutrino sources
this is negligibly small compared to the DM event rate, we are considering here, cf.\ eq.~\eqref{eq:Rate}.

The differential event rate of air molecules hitting the target is
\begin{equation}
 \frac{\diff N}{\diff t \diff v} = n \, A \, |v| \, f(v)  \,
\end{equation}
where $A$ is the surface of the target, $v$ the velocity of
the air molecules in the direction probed by the experiment,
$f(v)$ its standard Maxwellian distribution and $n = P/(k_B T)$
is the number density of the molecules with $P$ the pressure,
$k_B$ the Boltzmann constant and $T$ the temperature.
Note that the target is assumed to be symmetric and can be hit from
both sides. The hit rate after integrating over all velocities is hence,
\begin{align}
\frac{\diff N}{\diff t} &= \frac{P}{k_B T} A \sqrt{ \frac{2 k_B T}{\pi \, m} } = P \, A \sqrt{\frac{2 }{\pi \, m \, k_B T}} \\
 &\approx 8.3 \times 10^9 \left( \frac{P}{10^{-10} \text{ mbar}} \sqrt{\frac{\text{20 K}}{T}} \frac{A}{\text{mm}^2}   \right)   \frac{1}{\text{s}} \;.
\end{align}
Here we have assumed that the residual gas consists of molecular
hydrogen which is usually dominant in ultra-high vacuum,
cf.~e.g.~\cite{TheLIGOScientific:2014jea}, since vacuum pumps
can remove heavier molecules more efficiently.
For the pressure, we assume a vacuum system like in the
LHC~\cite{CERN-Brochure-2017-002-Eng} or advanced LIGO~\cite{TheLIGOScientific:2014jea}
and for the temperature
a cryonic system as in KAGRA~\cite{Somiya:2011np}.
The target surface is assumed to be 1~mm$^2$.
We want to remind the reader, that lower pressures and temperatures
have been achieved experimentally and there can be room for improvement,
but as a conservative estimate we use numbers
from current similar experiments if possible.

Now that poses a problem at first, since this rate is much larger than
a realistic DM hit rate. Let us assume in the following that the data bins
are $\delta t = 0.1$~ns long, which is common in optical experiments.
Then we expect about one hit from residual molecules in one data bin.

Our approach is still feasible, in principle, since the distribution of recoil momenta from
DM and the atmosphere are different. For this let us first calculate the recoil
momentum distribution per data bin. The differential recoil momentum
is
\begin{equation}
 \frac{\diff q_{\text{atm}}}{\diff t} = q \frac{\diff N}{\diff t} = 2 \, m \, n \, A \, v \, |v| \, f(v) \diff v \;,
\end{equation}
where we assume that each molecule just bounces off the target, $q = 2 \, m \, v$.
The expectation value is obviously zero (we neglect any potential airflow in the chamber).
The variance of the Gaussian distribution is then
\begin{align}
 \sigma_{q_{\text{atm}}}^2 &= \langle q_{\text{atm}}^2 \rangle = 4 \, m^2 n \, A \, \delta t \, \int v^2 |v| f(v) \diff v \\
  & = 4 \, m^2 n \, A \, \delta t \left[ \sqrt{ \frac{8}{\pi} }  \left( \frac{k_B T}{m} \right)^{3/2} \right]
      = 8 \sqrt{ \frac{2}{\pi} }  P \, A \, \delta t  \sqrt{m \, k_B T} \\
    &\approx 6.1 \times 10^{-48}  \left( \frac{P}{10^{-10} \text{ mbar}} \frac{A}{\text{mm}^2} \frac{\delta t}{\text{0.1 ns}}   \sqrt{ \frac{T}{\text{20 K}} }  \right) \frac{\text{kg}^2 \text{m}^2}{\text{s}^2} \;.   
\end{align}
From data bin to data bin we hence expect a momentum uncertainty
of order of $\sigma_{q_{\text{atm}}} \approx 2.5 \times 10^{-24}$~kg~m/s.
A cut on the recoil momentum can therefore suppress the fluctuations induced by
the residual gas hits significantly and we can still hope to see a DM signal as long as
the DM recoil momentum spectrum allows for larger recoil momenta.
In every realistic experiment, we would expect anyway  such minimal resolution of
the recoil momentum and that brings us
to the second of the experimental uncertainties.

The second major deviation from an ideal experiment is the uncertainty of the position
and/or the recoil momentum measurement. Considering a harmonic
oscillator with an eigenfrequency $\omega_0$ and a position resolution $d_{\text{min}}$
as a classical toy model, this
implies a minimal momentum resolution of the order of
$q_{\text{min}} = M_T \, \omega_0 \, d_{\text{min}}$.
To improve the sensitivity one could imagine to split the detector mass into
smaller detector cells, which would lower $q_{\text{min}}$ linearly with the number of cells.
As sensitivity goal we use a rather naive interpretation of the LIGO resolution
$d_{\text{min}} = 10^{-19}$~m at a frequency
of $\omega_0 = 100$~Hz \cite{TheLIGOScientific:2016agk}
which leads to a minimal momentum resolution of the order of
$\mathcal{O}(10^{-23})$~kg\,m/s. For definiteness we will use in the following
$q_{\text{min}} = 2 \times 10^{-23}$~kg\,m/s unless specified otherwise.
This number corresponds to the maximum recoil momentum for DM with a mass of a
few MeV with an order one uncertainty from the velocity distribution so that we can expect
to potentially see a signal for DM masses above that. Furthermore, with this
choice the background from the residual air molecules is very efficiently suppressed
and we end up with an effective rate
\begin{equation}
R_{\text{atm}}^{\text{cut}} \approx 5 \times 10^{-6} \text{ Hz} \;.
\end{equation}
Applying this cut makes the experiment almost free from that noise.

LIGO with its 40~kg mirrors has a much larger momentum cutoff
and cannot resolve the momentum recoil of an individual light DM hit
or the many hits of air molecules.
One might also wonder, if a similar resolution could be achieved anytime soon
in an experiment with much lighter mirrors as we require.
For instance, the so-called standard quantum limit in gravitational wave detectors \cite{Hawking:1987en}
is proportional to $1/(M_T \, \omega^2 \, L^2)$, where $L$ is the arm length.
Lighter target masses could hence be a problem. But looking at membranes with
eigenfrequencies in the kHz- or even MHz-range excited by DM hits instead,
cf.~\cite{Tsuchida:2019hhc} can compensate for that lighter masses. A more
thorough study of such matters will be
presented elsewhere \cite{Lee:2020xx}.
Nevertheless, the effect of DBM is independent of laser interferometers.
And as we had stated at the beginning of this section already the numbers
here should be seen as a target sensitivity for which our 
suggested method could become competitive to conventional direct DM searches,
cf.\ Fig.~\ref{fig:CrossSectionVsDMmass} and even reach lighter DM masses.

From that point of view our suggestion should be much more sensitive to heavier DM
since the expected recoil momentum would be larger. Nevertheless, current constraints
on the DM cross section to ordinary matter and the local DM number density suggest that
our approach is not competitive to conventional experiments in that mass range.

The uncertainty
in the momentum measurement affects recoil momenta larger than $q_{\text{min}}$ as well,
but in the asymmetry $A$ only the sign of the momentum enters which we know well,
after we discarded
all events with recoil momenta smaller than $q_{\text{min}}$.

A momentum cutoff $q_{\text{min}} > 0$ on one hand reduces the number of events in the signal
region and thus potentially lowers the significance. On the other hand, it can
enhance $A$ and the two effects are competing with each other.

In Fig.~\ref{fig:CrossSectionVsDMmass} we show the final estimated sensitivity
curve including the noise from residual air molecules and the momentum resolution
cutoff as the red line given by the condition
\begin{equation}
\frac{\langle A \rangle^2}{\sigma_A^2 + \sigma_{\text{bkg}}^2}  = 4 \;,
\end{equation}
where $\langle A \rangle$ and $\sigma_A$ include both of the DM signal
and the background while $\sigma_{\text{bkg}}$ only includes the hits from the residual gas. 

First of all, we can see in
Fig.~\ref{fig:CrossSectionVsDMmass} that the momentum cutoff makes it impossible
for our hypothetical experiment to discover DM masses less than $\mathcal{O}(10)$~MeV/c$^2$
(the vertical dashed line is at 9.3~MeV/c$^2$).
This will be clear from the kinematical arguments which we discuss later.
But we can also see that for the mass range above 15~MeV/c$^2$
the momentum cutoff
increases the theoretical sensitivity compared to the idealized experiment
which is due to the increase in the asymmetry and the very efficient background rejection.

From Fig.~\ref{fig:CrossSectionVsDMmass} it is also clear, that the $q_{\text{min}}$
plays the equivalent role of the energy threshold in conventional DM detectors.
In our case though, this depends only indirectly on the chosen target material, instead it
depends dominantly on the resolution of the interferometer and the target mass neglecting
potential excitation modes of the target material.

To give an impression on how our sensitivity compares to experimental results, we also show
the exclusion lines from 
CRESST \cite{Angloher:2017sxg, Abdelhameed:2019hmk}, EDELWEISS (Migdal) \cite{Armengaud:2019kfj}
and CDEX (Migdal) \cite{Liu:2019kzq} which are
relevant for DM masses above 50~MeV/c$^2$. Cosmology also provides bounds, but they are rather weak. Refs.~\cite{Xu:2018efh, Nadler:2019zrb},
for instance,
find that the DM-nucleon cross section over the whole displayed mass region
should be less than $\mathcal{O}(10^{-28})$~cm$^2$.

\begin{figure}
	\centering
	\includegraphics[width=0.7\textwidth]{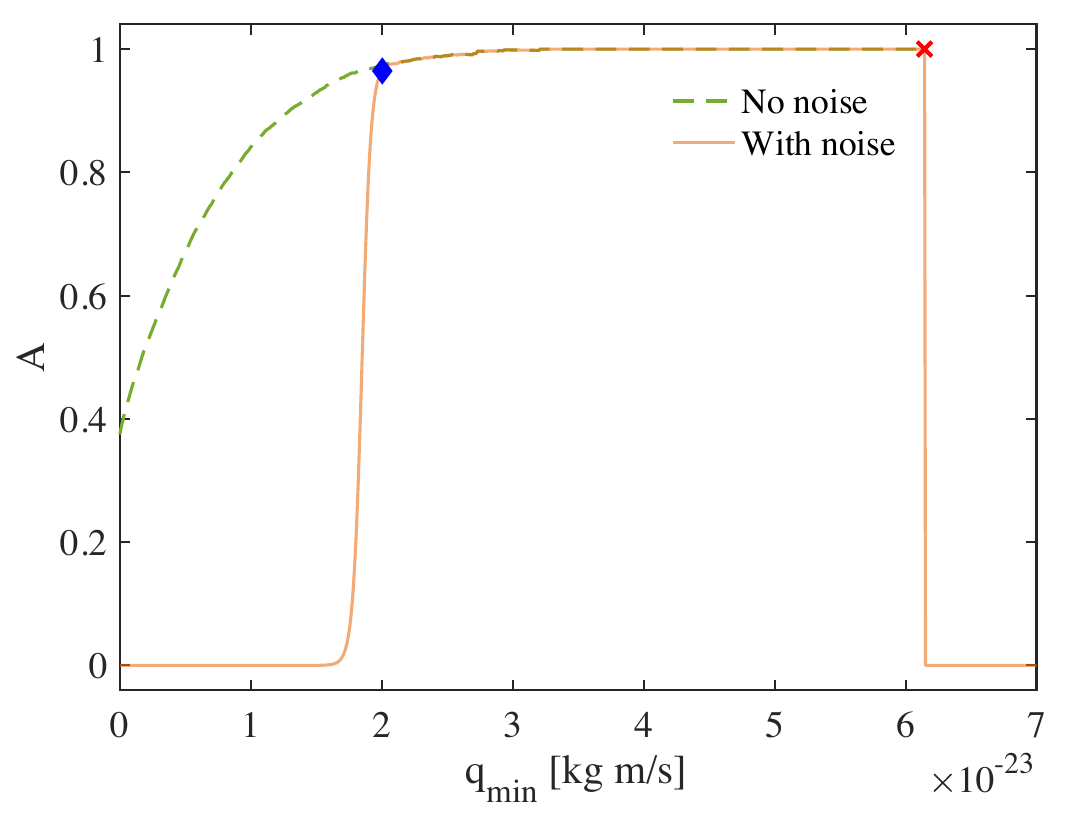}%
	\caption{
		The dependence of the asymmetry $A$
		on a lower bound for the recoil momentum $q_{\text{min}}$
		for a benchmark point with (solid line) and without noise from residual gas (dashed line).
                 As soon as $q_{\text{min}}$ is larger than
		the maximal DM recoil momentum there are no events left and the lines end at the red cross.
		The blue diamond corresponds to the blue diamond in Fig.~\ref{fig:CrossSectionVsDMmass}.
		For more explanations, see main text.
		\label{fig:Avsq}
	}
\end{figure}

It is instructive to have a closer look at the momentum cutoff $q_{\text{min}}$.
We can easily understand from the most extremal case that the DM asymmetry $A$ can increase for
an increasing cutoff.
Let $v_{\text{lab}}$ be the lab velocity with respect to the galactic halo in
the direction we are probing.
For $|\vesc - v_{\text{lab}}| < q_{\text{min}}/(2 \, \mdm) < |\vesc + v_{\text{lab}}|$
the asymmetry is extremal, $|A| = 1$, due to simple kinematics
and ignoring any potential backgrounds.
This leads to another characteristic feature of a DM signal. Increasing
$q_{\text{min}}$ the modulus of the DM induced asymmetry continuously
increases until no events are left for an idealized experiment.
This is the dashed green line in Fig.~\ref{fig:Avsq},
where we have set
$\sigma_{\text{DM}-N} = 2.2 \times 10^{-33}$~cm$^2$, $\text{M}_{\text{DM}}$ = 22 MeV/c$^2$
which ends at $q_{\text{min}} = 6.1 \times 10^{-23}$~kg\,m/s.
There is of course a trade-off. A larger $q_{\text{min}}$ leads to a larger DM asymmetry,
but it will also lead to a decreased DM event rate.

This will additionally be affected by the background
from residual gas. We make here the same assumptions as
we have used in Fig.~\ref{fig:CrossSectionVsDMmass}
and for
the cross section and the dark matter mass we use the same numbers as for
the idealized experiment here. From the solid orange line in Fig.~\ref{fig:Avsq}
we see that a very small $q_{\text{min}}$ the observed asymmetry would drop to zero
since the number of background events is many orders of magnitude larger than the DM
events pushing the asymmetry to zero. We also see that the threshold for which this background
becomes small is very sharp since the background distribution is comparatively narrow compared
to the DM recoil momentum distribution. 
Note that in theory we still expect an extremely tiny number of background events from the very tail
of the distribution such that $A$ drops again to zero above the maximum DM recoil momentum.
Varying $q_{\text{min}}$ in the data analysis hence might
reveal a previously unidentified background or signal, if it exhibits such features.

Although in our simplified sketch setup we did not consider that possibility explicitly there
might be backgrounds which have a broader recoil momentum distribution than the potential
DM signal leading to another source of events above the DM cutoff.
That could be cosmic rays or radioactive decays in the chamber, which are difficult
to model realistically without specifying the location, materials, shielding, etc. which goes beyond
the scope of this paper.
Generally speaking this would reduce the sensitivity, shifting the
sensitivity curve upwards since the DM asymmetry would be washed out.
To reduce such backgrounds, one can introduce an upper bound on
the recoil momentum $q_{\text{max}}$ in the data analysis.

A careful study of the dependence of the measured asymmetry $A$ on both,
$q_{\text{min}}$ and $q_{\text{max}}$, hence might help to understand the background
much better in the absence of a signal.
Should a potential signal be identified the variation of both
would provide valuable information on the recoil momentum spectrum
of background and signal.
In combination with the
information on the time dependence that would be very exciting and would
allow to constrain the DM mass scale and halo model.

\section{Summary and Conclusions}

In this paper we have discussed the effect of
Dark Matter induced Brownian motion which could be used as
a novel way to search for light DM.
This motion has a characteristic, time- and direction-dependent asymmetry of the recoil
momenta due
to the relative motion of the earth to the DM halo. This asymmetry
and, in particular, its daily modulation
could in principle
be observed in an experiment, which measures
the position/velocity of a target object to very high precision.
Currently, interferometers seem to be most promising
for that purpose, but not LIGO and other earth-bound telescopes
since their mirror masses
imply a momentum cutoff well above our expectations for light DM.
In the future though, gravitational wave detectors 
might be able to detect vibrational modes of thin mirrors excited
by DM scatterings \cite{Tsuchida:2019hhc}.

Here we have assumed a hypothetical, different kind of 
experiment (which does not even have to be based on laser interferometers)
with lighter target masses. If this experiment can resolve recoil momenta
as small as discussed in our paper
one could test DM masses down to the $\mathcal{O}(10)$~MeV/c$^2$ region.
For our example, we find that for a DM mass of 22~MeV/c$^2$ one could test
a DM-nucleon cross section
down to $2.2 \times 10^{-33}$~cm$^2$ with only 10 minutes of data taking.
Further research regarding the feasibility of this numbers considering
different experimental approaches is needed in the future.

Our results also depend on other uncertainties like
the DM velocity distribution and modifications
of that by cosmic ray scatterings or local streams might enhance or weaken
considerably the sensitivity of such an experiment.

\appendix
\subsection*{Acknowledgments}
We would like to thank Yue-Lin Sming Tsai and Ulises Salda$\tilde{\text{n}}$a-Salazar for useful comments on the manuscript.
We also thank the EPJ C referee for their valuable suggestions and comments.
TC and MS are supported by the Ministry of Science and Technology (MOST) of Taiwan 
under grant number MOST 107-2112-M-007-031-MY3.

\end{document}